# Modeling Electrical Switching of Nonvolatile Phase-Change Integrated Nanophotonic Structures with Graphene Heaters

*Jiajiu Zheng,[†] Shifeng Zhu,[‡] Peipeng Xu,[⊥] Scott Dunham,[†,‡] Arka Majumdar\*[,†,‡]*

[†]Department of Electrical and Computer Engineering, University of Washington, Seattle, WA 98195, USA

[‡]Department of Physics, University of Washington, Seattle, WA 98195, USA

[⊥]Laboratory of Infrared Materials and Devices, Advanced Technology Research Institute, Ningbo University, Ningbo 315211, China

\*E-mail: arka@uw.edu

ABSTRACT: Progress in integrated nanophotonics has enabled large-scale programmable photonic integrated circuits (PICs) for general-purpose electronic-photonic systems on a chip. Relying on the weak, volatile thermo-optic or electro-optic effects, such systems usually exhibit limited reconfigurability along with high energy consumption and large footprints. These challenges can be addressed by resorting to chalcogenide phase-change materials (PCMs) such as $Ge_2Sb_2Te_5$ (GST) that provide substantial optical contrast in a self-holding fashion upon phase transitions. However, current PCM-based integrated photonic applications are limited to single devices or simple PICs due to the poor scalability of the optical or electrical self-heating actuation approaches. Thermal-conduction heating via external electrical heaters, instead, allows large-scale integration and large-area switching, but fast and energy-efficient electrical control is yet to show. Here, we model electrical switching of GST-clad integrated nanophotonic structures with graphene heaters based on the programmable GST-on-silicon platform. Thanks to the ultra-low heat capacity and high in-plane thermal conductivity of graphene, the proposed structures exhibit a high switching speed of ~80 MHz and high energy efficiency of 19.2 aJ/nm$^3$ (6.6 aJ/nm$^3$) for crystallization (amorphization) while achieving complete phase transitions to ensure strong attenuation (~6.46 dB/μm) and optical phase (~0.28 $\pi$/μm at 1550 nm) modulation. Compared with indium tin oxide and silicon p-i-n heaters, the structures with graphene heaters display two orders of magnitude higher figure of merits for heating and overall performance. Our work facilitates the analysis and understanding of the thermal-conduction heating-enabled phase transitions on PICs and supports the development of the future large-scale PCM-based electronic-photonic systems.

The past decades have witnessed the booming applications of photonic integrated circuits (PICs). Benefiting from the low-loss broadband transmission, PICs have demonstrated advantages over electronics in information transport including telecommunication and data center interconnects. Recently, thanks to the remarkable advances in nanofabrication, the level of complexity of photonic integration has reached a new height, shedding light on the future electronic-photonic systems on a chip.[1-2] The availability of large-scale PICs, along with the slowing down of Moore's Law[3] and the von Neumann bottleneck in electronics, is thus offering PICs new opportunity to compete with electronic systems in energy-efficient broadband data processing and storage, in particular, for emerging applications such as neuromorphic computing,[4] quantum information,[2,5] and microwave photonics.[6-7] Similar to electronic field-programmable gate arrays (FPGAs), success in these fields requires large-scale programmable PICs that have low-energy, compact, and high-speed building blocks with ultra-low insertion loss.[8-9] Such general-purpose PICs can be reconfigured at will to meet the need for specific applications. Whereas numerous programmable photonic systems have been reported,[4-7,10] limited tunability of the systems is exhibited due to the weak thermo-optic or electro-optic effects of the materials, leading to high energy consumption and large chip footprints. Microelectromechanical systems (MEMS)[11] or resonator-based systems[12-13] can help improve the modulation strength, but they suffer from either a large actuation voltage up to 40 V or narrow optical bandwidth as well as high sensitivity to fabrication and temperature variations.[14] The volatile reconfigurability of these PICs also necessitates a continuous power supply rendering the systems energy-inefficient.

To address these challenges, it is highly desirable to explore other active photonic materials with strong optical modulation and self-holding characteristics. Chalcogenide phase-change materials (PCMs) such as $Ge_2Sb_2Te_5$ (GST) are one such candidate.[15-16] Firstly, PCMs can retain crystalline and amorphous states with long retention time (for at least 10 years)[17-18] and extremely high contrast in resistivity[17] and complex refractive index (usually $\Delta n > 1$) over a wide spectral range,[15-16,19] thus enabling ultra-compact,[20-25] broadband,[16,22,25] and multi-level[16,23-24,26-32] operations for nonvolatile integrated photonic applications without static energy consumption. Secondly, phase transitions of PCMs can be reversibly actuated by ultra-short (picosecond to nanosecond) optical or electrical pulses[33-35] with high cyclability (potentially up to $10^{15}$ switching cycles)[25,36] and low energy (down to femtojoules per bit or ~10 aJ/nm³).[16,37-39] Moreover, PCMs are highly scalable[40] and compatible with other substrates without the "lattice mismatch" issue as the as-deposited PCMs are usually in the amorphous state. Therefore, PCMs have been introduced to a variety of programmable PIC applications (usually placed on top of waveguides), including optical switches/modulators,[16,20-24,30,41-44] photonic memories,[26,29,31] and optical computing.[27-28,32,45-46] Recently, mixed-mode operations[39,47] and tunable volatility[48] of PCMs have also been demonstrated. However, current PCM-based applications are limited to single devices or simple PICs. To scale up the PCM-integrated photonic devices to a much higher complexity as required by the future photonic FPGAs, it is important to have scalable control over the states of PCMs.

In general, the phase transitions of PCMs on PICs can be triggered either by self-heating or thermal-conduction heating. Self-heating relies on the photothermal or Joule heating effect of PCMs to actuate the phase change process and can be realized by free-space optical switching,[16,20,41] on-chip optical switching,[26-32,38-39,43,46-49] or electrical threshold switching.[21,23,39,42] Resembling the approach used in rewritable optical disks, free-space optical switching, where PCMs are heated up by focusing laser pulses onto target devices in the far field, facilitates the switching of large-area PCMs at any position, but is not optimized for further integration and scaling due to the slow, diffraction-limited, inaccurate alignment process.[16] In contrast, on-chip optical switching, mainly exploiting the evanescent coupling of near-field light pulses between waveguides and above PCMs, allows fully integrated all-optical operations of small-size PCMs down to the nanoscale.[39] However, it is challenging to switch large-area PCMs through this method due to the nonuniform heating, and the complexity of PICs is restricted because of the difficulty in light routing and cascaded-device heating.[28] Note that, both the photothermal-based approaches will suffer from the low extinction coefficient of amorphous state in the crystallization process, where multiple pulses or a single structured pulse are usually needed. The issue becomes particularly severe when transparent PCMs are used.[50-51] Electrical threshold switching by contacting the two sides of PCMs in a circuit seems to be a better choice for large-scale integration. Nevertheless, it proves to be ineffective because the limited operation volume of PCMs due to the crystallization filamentation[52] and nonuniform heating conflicts with the requirement of relatively large size for photonic devices, resulting in low optical contrast between two states.[39] This method will also face challenges when it comes to less conductive PCMs which are generally also transparent due to their larger bandgaps. In comparison, thermal-conduction heating via external electrical heaters,[24-25,30,44] instead, can locally select and arbitrarily extend the switching region by increasing the size of the heaters. Therefore, this approach intrinsically eliminates all the above problems, enabling large-area phase-change photonic devices[22,25] with the potential of strong optical modulation and high-complexity integration. However, among the few works based on this approach,[24-25,30,44] no fast (> 10 MHz) and energy-efficient (~10 aJ/nm³) electrical control has been reported so far. To maximize the advantages of thermal-conduction heating in terms of both optical and heating performance, the optical waveguide and heating system of phase-change integrated nanophotonic cells (PINCs), *i.e.* the fundamental unit of PCM-integrated photonic devices, remain to be improved and the heating process requires to be analyzed and optimized. To assist the development of the future large-scale PCM-based PICs, it is also important to develop a comprehensive model that has the capability of controlling and predicting the device performance.

In this work, we propose and model electrical switching of nonvolatile PINCs with graphene heaters based on the programmable GST-on-silicon platform.[16,22,25] Thanks to the ultra-low heat capacity and high in-plane thermal conductivity of graphene,[53-54] the proposed structures exhibit a high switching speed of ~80 MHz and high energy efficiency of 19.2 aJ/nm³ (6.6 aJ/nm³) for crystallization (amorphization) while achieving complete phase transitions to enable high optical contrast (~6.46 dB/μm or ~0.28 π/μm at 1550 nm). Further analysis implies that gigahertz operations and energy efficiency near the fundamental limit[36] are possible for partial crystallization or amorphization that can be applied in multi-level operations. By comparing graphene with indium tin oxide (ITO) and silicon p-i-n diode heaters, we conclude that the PINCs with graphene heaters have the best heating and overall performance with two orders of magnitude higher figure of merits. By tuning the Fermi level ($E_F$) of graphene to the Pauli blocking region,[55-58] even better optical performance and lower operation voltage can be achieved.

RESULTS AND DISCUSSION

**Device Configuration and Modeling.** As illustrated by Figure 1a, the proposed PINC is composed of a GST-on-silicon hybrid waveguide (where a thin film of GST with a width of $w_{GST}$ and a thickness of $h_{GST}$ is placed on top of the silicon rib waveguide) with a certain length ($L$) based on a silicon-on-insulator (SOI) wafer with a 220-nm-thick silicon layer on top of a 3-μm-thick buried oxide. The geometry is similar to the one recently reported for optical switching.[16] To conduct the electrical switching, the PINC is conformally covered with monolayer graphene as the external heater with palladium contacts and a capping layer of $SiO_2$. Graphene, a single layer of carbon atoms arranged in a honeycomb lattice, has recently been introduced to integrated nanophotonic devices as a transparent heater[59-61] due to its high intrinsic in-plane thermal conductivity, ultra-low heat capacity, tunable transparency and conductivity as well as good flexibility and compatibility with complementary metal-oxide-semiconductor (CMOS) processes.[53-58,62] Consequently, graphene is a promising candidate for external heaters in PINCs with great potential for high-speed and low-energy electrical switching. In order to evaluate the optical performance of the PINC, the input and output ports of the PINC are assumed to be connected with regular silicon waveguides, which is the case for most applications. Optical mode analysis is performed based on the frequency-domain finite-element method (FEM) using COMSOL Multiphysics (see Methods).

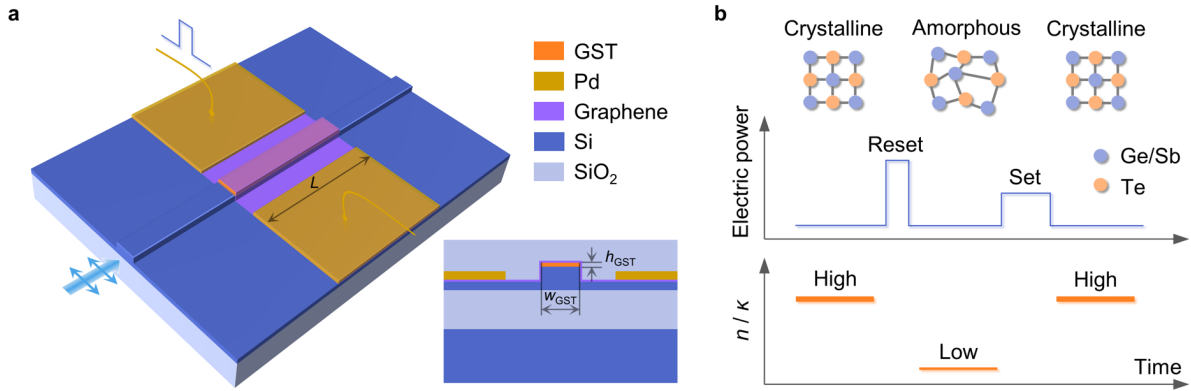

**Figure 1.** Device configuration. (a) Schematic of the proposed PINC with a graphene heater. Here, the silica cladding is hidden for clarity. The two ports of the PINC are connected with regular silicon rib waveguides in this case. Inset: cross-section of the hybrid cell. (b) Operation principle of electrical switching of the PINC with an external heater. The lattice structure, refractive index ($n$), and extinction coefficient ($\kappa$) of GST for amorphous and crystalline states and the electric power of the reset and set pulses are illustrated.

Differing from the self-heating approaches, electrical switching using external heaters relies on the transfer of electrical pulse-generated Joule heat from heaters to PCMs to actuate the phase transitions. For amorphization (Figure 1b, Reset), a single pulse with high power is applied to the contacts to increase the temperature of PCMs above the melting point ($T_m$) and then immediately removed to obtain rapid quenching, leaving PCMs in the disordered glass state with low refractive index ($n$) and extinction coefficient ($\kappa$). For crystallization (Figure 1b, Set), PCMs are heated just above the glass transition temperature ($T_g$) but below $T_m$ by applying a pulse with relatively low power for a long time to enable nucleation of small crystallites and subsequent growth of them, resulting in high optical constants in the crystalline state. To analyze the Set and Reset processes of the proposed PINC, we have established a fully coupled electro-thermal two-dimensional (2D) time-domain FEM model (see Methods) based on the cross-section of the PINC (Figure 1a, inset). Here, we assume that the phase transitions smoothly occur in a small temperature interval of $\Delta T_m$ = 10 K ($\Delta T_g$ = 100 K) centered at $T_m$ = 888 K ($T_g$ = 673 K) for the fast melting (the relatively slow crystallization).[52,63] Our model can be further improved by incorporating accurate kinetic description of the phase transitions, including melting, vitrification, nucleation, and growth. 2D simulations are more than sufficient to evaluate the heating performance given that the geometry of the PINC remains constant in the light propagation direction (assumed to be 1 μm in this work) and the boundary-effect-induced nonuniform heating in the interface between the PINC and the regular waveguides can be alleviated by fully covering the PINC with a larger heater.[44]

**Optical Performance.** For large-scale programmable photonic applications, strong optical modulation and low insertion loss are essential for the optimal optical performance of the PINCs. Here, we define an optical figure of merit $FOM_1 = \Delta n_{eff}/\kappa_{effa}$, where $\Delta n_{eff}$ denotes the effective refractive index ($n_{eff}$) change between the PINC with crystalline GST (cGST) and amorphous GST (aGST) that determines the modulation strength and $\kappa_{effa}$ is the effective extinction coefficient ($\kappa_{eff}$) of the PINC with aGST that reflects the insertion loss of the device as generally the loss for cGST is much larger than that for aGST. Both of the parameters are of great importance for optical phase modulation and the previously reported phase-change coupling modulation.[22,64] One can also define the optical figure of merit as $\Delta \kappa_{eff}/\kappa_{effa}$ for attenuation modulation (with $\Delta \kappa_{eff}$ being the effective extinction coefficient change). However, here we will primarily focus on $FOM_1$ since both of the figure of merits have similar behavior (Figure S1). As the optical performance of the GST-on-silicon waveguides are quite broadband near 1550 nm for at least 40 nm,[16] the following analysis is conducted at the single wavelength of $\lambda$ = 1550 nm.

Improvement of $FOM_1$ can be achieved by optimizing the geometry of the GST-on-silicon hybrid waveguide. As $FOM_1$ does not strongly depend on the dimensions of the silicon rib (Figures

S1a, S1b), only the influence of the size of the GST film is discussed here while the width and height of the silicon rib are fixed to be 500 nm and 120 nm, respectively. According to the mode analysis, strong modification of the mode profile and the complex effective index ($\tilde{n}_{\text{eff}} = n_{\text{eff}} - \kappa_{\text{eff}}$ i) can be observed once the GST is electrically switched between the amorphous (Figure 2a) and crystalline (Figure 2b) states, indicating substantial refractive and absorptive modulation effects. Figures 2c and 2d summarize the variation of the effective refractive index and attenuation coefficient ($\alpha = 4\pi\kappa_{\text{eff}}/\lambda$) of the hybrid waveguide with respect to the GST geometry for aGST and cGST. As expected, both the parameters and $FOM_1$ (Figures S1c, S1d) increase with the increase of the GST width and thickness. Therefore, in the following analysis, we select the width of the GST film to be as large as 500 nm (same as the width of the silicon rib). However, the thickness of the GST film is set to be 20 nm by default unless otherwise specified considering the trade-off between $FOM_1$ and the difficulty in switching (to be discussed later) as well as the signal-to-noise ratio in real experiments.

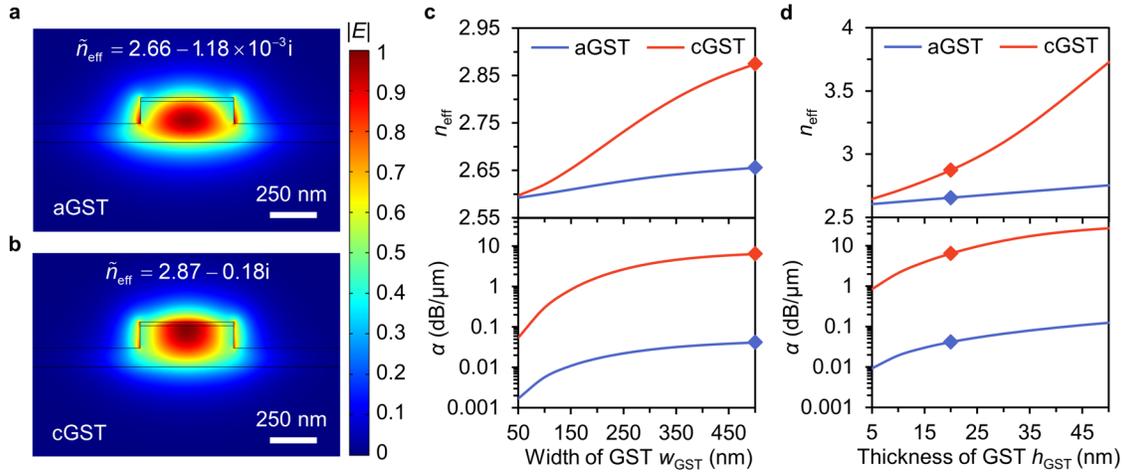

**Figure 2.** Dependence of optical performance on the waveguide geometry. (a, b) Normalized electrical field profile and complex effective index ($\tilde{n}_{\text{eff}}$) of the fundamental quasi-transversal electric (TE) mode of the GST-on-silicon hybrid waveguide with (a) aGST and (b) cGST at a wavelength ($\lambda$) of 1550 nm. The width and thickness of the GST are 500 nm and 20 nm, respectively. (c, d) Effective refractive index ($n_{\text{eff}}$) and attenuation coefficient ($\alpha = 4\pi\kappa_{\text{eff}}/\lambda$, $\lambda = 1550$ nm) of the hybrid waveguide as a function of the (c) width and (d) thickness of the GST with $h_{\text{GST}} = 20$ nm in (c) and $w_{\text{GST}} = 500$ nm in (d). The marked dots correspond to the structure in (a) and (b) and denote the adopted waveguide geometry for the following analysis unless specifically pointed out. Here, the width and height of the silicon rib are fixed to be 500 nm and 120 nm, respectively. Neither heaters nor metal contacts are involved at this stage.

The optical performance can also be enhanced by reducing the additional loss from the metal contacts and heaters. Although placing the electrodes far away from the rib can theoretically avoid high insertion loss (Figure S2), it will compromise the energy efficiency since part of the generated heat will be dissipated around the long slab region. In this case, we choose the distance between the electrodes and the rib to be as close as 500 nm (the thickness of the metal is set to be 50 nm) while maintaining moderate additional loss of ~0.01 dB/μm. The wet-transferred graphene grown by chemical vapor deposition usually has a Fermi level of around $-0.28$ eV $\sim -0.23$ eV.[65-66] Here, we assume the Fermi level of graphene to be $-0.26$ eV that leads to additional loss of ~0.1 dB/μm (see Methods and Supporting Section 6). To suppress the loss from graphene, one can electrically tune the Fermi level of graphene to the Pauli blocking region where interband transitions of electrons are prohibited (*i.e.* $E_F < -0.4$ eV for the wavelength of 1550 nm) through a gate electrode.[55-58] This can potentially reduce the loss of graphene to ~0.002 dB/μm (see Methods and Supporting Section 6) while the increased energy consumption is significantly less than the switching energy (to be discussed later) and thus is negligible. As a result, our proposed PINC exhibits a $FOM_1$ of ~50 (~140 with gated graphene) and propagation loss per unit length of ~0.15 dB/μm (~0.05 dB/μm with gated graphene) with an attenuation modulation of ~6.46 dB/μm and an optical phase modulation of ~0.28 π/μm at 1550 nm. The loss due to the mode mismatch between the regular silicon waveguide and the hybrid waveguide is ~0.03 dB on each side.

**Heating Performance.** In order to operate the electrical switching with high heating performance in terms of high switching speed and energy efficiency, the real-time temperature ($T$) distribution of the PINC in response to an electrical pulse is calculated and analyzed based on the electro-thermal model. To successfully actuate the phase transitions without damaging the device, the raised temperature during the heating must be subjected to several constraints (the cooling rate during the quenching is also required to be about $10^{10}$-$10^{11}$ K/s[52] that is generally satisfied in our simulations). For crystallization, the temperature within the GST should be greater than $T_g + \Delta T_g/2$ to ensure adequate nucleation and growth but less than $T_m - \Delta T_m/2$ to prevent re-amorphization. For amorphization, the temperature of the GST has to be elevated above $T_m + \Delta T_m/2$ but not so high to induce ablation. Besides, the temperature within the electrodes, heater, silicon waveguide, and silica cladding should always be kept below their melting points. In other words, the temperature gradient within the GST and the heat accumulating in other regions limit the implementation of the phase transitions.

Figure 3a presents a typical temperature profile (without phase transitions) of the PINC with aGST at the end of a pulse with electrical power ($P_0$) of 5 mW and a pulse width ($\Delta t$) of 50 ns. In particular, the data cut along the $x$ axis (Figure 3b) and $y$ axis (Figure 3c) reveal that due to the flatness of the graphene, the GST is almost uniformly heated in the horizontal direction except the edges. However, the temperature of the GST in the vertical direction exhibits a large gradient with a much higher value close to the heater. Therefore, the temperature gradient within the GST ($\Delta T$) can be represented by the absolute temperature difference between the top ($T_{top}$) and bottom ($T_{bottom}$) surface of the GST film along the $y$ axis (*i.e.* $\Delta T = |T_{top} - T_{bottom}|$). With a moderate temperature gradient (Figure S3, $\Delta T$ remains ~100 K during the phase transition), the crystallization process (Figure 3d, $P_0 = 14$ mW) can proceed without re-amorphization. However, a very high-power pulse or a thick GST film will result in serious re-amorphization (Figure S4) due to the large temperature gradient within the GST. Further analysis (Figures 3e and 3f) confirms that the temperature gradient at the end of a heating pulse (without phase transitions) rises dramatically with the increase of the pulse power and the thickness of the GST film but increases less sensitively with the pulse width. Since small $\Delta T$ is desirable for a practical Set (Reset) process without any re-amorphization (ablation) and melting of other materials, a thin film of GST and a pulse with moderate power are preferred. In the following analysis, we will investigate the influence of pulse power, pulse width, and the thickness of the GST film on the switching speed and energy efficiency.

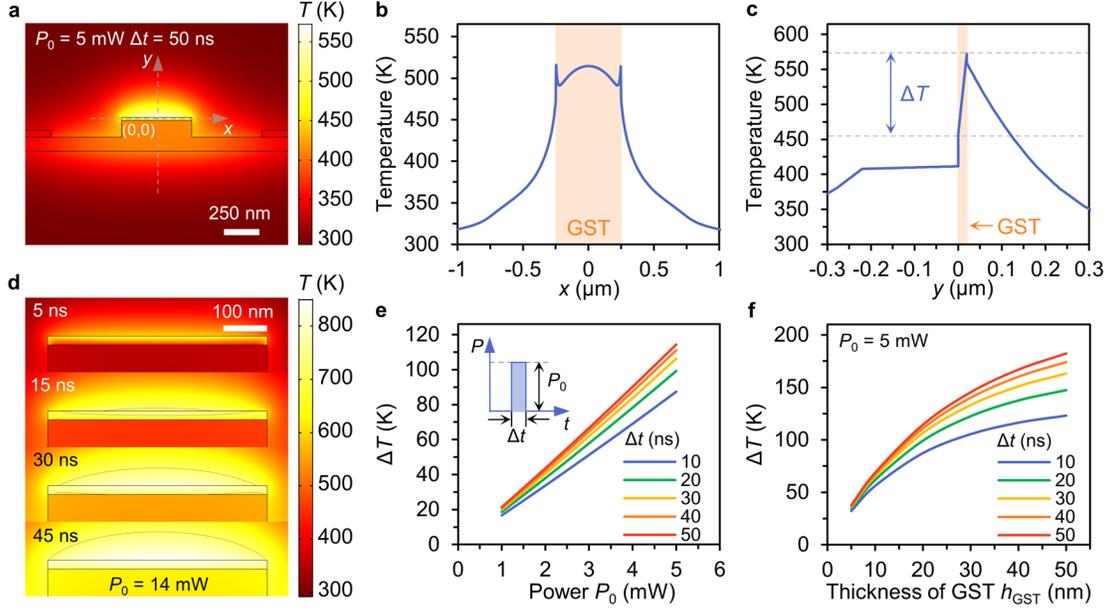

**Figure 3.** Temperature distribution analysis. (a) Temperature ($T$) profile of the PINC for aGST at the end of a pulse ($P_0$ = 5 mW, $\Delta t$ = 50 ns). No phase transition is induced due to this pulse. The dashed lines denote the coordinate system with the $x$ ($y$) axis parallel (perpendicular) to the surface of the GST film and the origin located at the center of the GST cross-section. (b) Temperature profile along the $x$ axis in (a). (c) Temperature profile along the $y$ axis in (a). The orange shaded areas in (b) and (c) represent the position of the GST film. (d) Temperature distribution at different times for a crystallization (Set) process ($P_0$ = 14 mW). The red lines denote the contour of $T$ = 723 K, within which the GST is assumed to be crystallized. (e) Temperature gradient ($\Delta T$, also marked in (c)) at the end of a pulse as a function of pulse power for different pulse widths. Inset: illustration of the applied pulse. (f) Temperature gradient at the end of a pulse as a function of GST thickness for different pulse widths ($P_0$ = 5 mW). The pulse energy is selected to be sufficiently low to avoid causing any phase transition of aGST in (e) and (f).

The switching speed of the PINC is limited by the pulse width and the dead time ($\tau$, 1/e cooling time) due to the thermal relaxation. As the Set process usually requires a relatively longer pulse and thus determines the ultimate speed, we mainly discuss the thermal relaxation of cGST and the crystallization period ($t_{ac}$, defined as the summation of the required pulse width $\Delta t_{ac}$ and corresponding dead time $\tau_{ac}$ for crystallization). As presented in Figures 4a and 4b, the transient temperature response due to the heating and cooling of the cGST (without phase transitions) shows considerably higher cooling rates for shorter pulses. This could be intuitively understood that for longer pulses, more energy will get lost into the waveguide and substrate due to the thermal diffusion,[26,29] so that a larger heat capacity and a longer thermal time constant are expected leading to a longer dead time (and a lower energy efficiency as will be discussed later). In contrast, the dead time depends little on the pulse power (Figure S5a) and the thickness of the GST film (Figures S6a, S6b). This is, however, not the case for the crystallization. The minimum pulse width ($\Delta t_{ac}$) required to achieve complete crystallization rapidly decreases with the increase of the pulse power (Figure S5b, with a fixed GST thickness) and linearly drops with the decrease of the GST thickness (Figure S6c, under fixed pulse power). As a result, the dead time ($\tau_{ac}$) decreases a lot accordingly (because the pulse width is shorter). This is especially conspicuous if we consider the optimal (fastest) case (Figure 4c) that the maximum pulse power allowed to actuate crystallization without re-amorphization (limited by the temperature gradient as discussed earlier) is applied. The crystallization period substantially reduces with the increased power for the thinner GST film. Therefore, to obtain high-speed operations, a thin film of GST and a short pulse (enabled by using high-power pulse) are needed. For the PINC with 20-nm-thick GST, the switching speed can be as fast as ~80 MHz (with $\Delta t_{ac}$ = 8 ns, $\tau_{ac}$ = 4.47 ns) that is one or more orders of magnitude larger than previous results.[24-25,30,44] It is worth noting that for partial crystallization, $\Delta t_{ac}$ can be even less than 1 ns for 10-nm-thick GST (Inset of Figure 4c), inferring that gigahertz operations are possible.

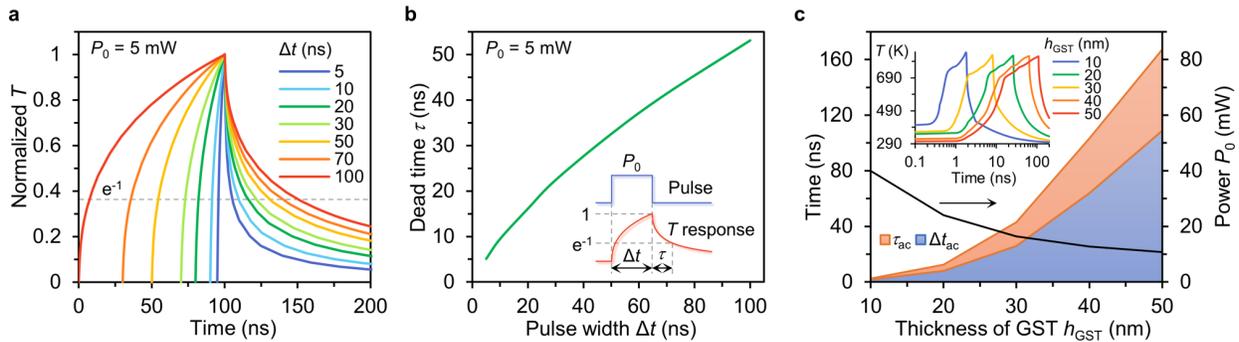

**Figure 4.** Transient response and speed analysis. (a) Normalized temperature response in the center of the GST cross-section to the pulses with different pulse widths. The power of the pulse is chosen to be as low as 5 mW in order to heat the cGST without inducing any phase transition. The dashed line denotes the position where the temperature is 1/e of the maximum. (b) Extracted dead time ($\tau$, 1/e cooling time) from (a) as a function of pulse width. Inset: illustration of the applied pulse and temperature response. (c) Area chart of the minimum pulse width ($\Delta t_{ac}$) and corresponding dead time ($\tau_{ac}$) required to achieve complete crystallization actuated by the maximum allowed pulse power (black line) as a function of GST thickness. Inset: transient temperature response of the crystallization process for different thicknesses of GST.

In order to achieve high energy efficiency, similar rules of thumb can be found. Here, the energy efficiency ($\eta$) is defined as the ratio of the absorbed heat energy in GST ($E_{\mathrm{GST}}$) at the end of a pulse and the applied electrical pulse energy ($E_{\mathrm{pulse}}$) and can be given by

$$\eta = \frac{E_{\mathrm{GST}}}{E_{\mathrm{pulse}}} = \frac{\int \rho C_p \left(T - T_0\right)\mathrm{d}V}{P_0 \Delta t} \tag{1}$$

where $\varrho$ is the material density, $C_p$ is the specific heat, $T_0$ is the initial ambient temperature (293 K), and the integral domain is over the entire GST film. As the Set process usually consumes more energy, we primarily discuss the energy efficiency of heating the aGST and crystallization. Similar to the dead time, the energy efficiency of heating (without phase transitions) significantly diminishes with the increase of the pulse width (Figure 5a) but is insensitive to the change of the pulse power (Inset of Figure 5a and Figure S7). However, the energy efficiency is improved with the increase of the GST film (Figure 5b). This may mislead one to choose a thick GST film for low energy consumption. Indeed, if the crystallization is involved, a thin film GST requires a much shorter pulse to optimally actuate the phase transitions (using the maximum allowed power, Figure 4c) that ultimately results in higher energy efficiency for crystallization ($\eta_{\mathrm{ac}}$, blue line in Figure 5c). In comparison, if a pump pulse with a fixed power is utilized, since the required pulse width does not increase much (Figure S6c), $\eta_{\mathrm{ac}}$ will still rise with the increase of the GST thickness (red line in Figure 5c). Consequently, a thin film of GST and a short pulse optimized by high power are critical to enable high energy efficiency. In other words, a fast PINC is also an energy-efficient device. For the PINC with 20-nm-thick GST, the consumed energy for crystallization can be optimized to be as low as ~0.192 nJ (19.2 aJ/nm³) with energy efficiency of ~3.5% that is at least one order of magnitude more efficient than previous results.[24-25,30,44] Based on a similar trend, the consumed energy for amorphization can be as low as 0.066 nJ (6.6 aJ/nm³) that is almost two or more orders of magnitude more efficient than previous results.[24-25,30,44] Note that the energy efficiency can be further improved if partial crystallization or amorphization is needed which is essential for multi-level operations. In this case, the pulse width can be much shorter (< 1 ns as discussed earlier) so that it is possible to reduce the energy consumption to near the fundamental limit (1.2 aJ/nm³).[36]

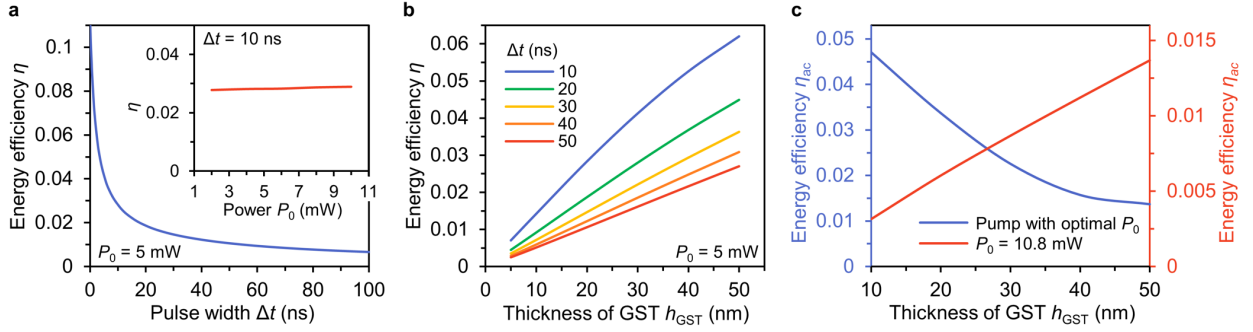

**Figure 5.** Energy efficiency analysis. (a) Energy efficiency ($\eta$) as a function of pulse width. Inset: pulse-power insensitive energy efficiency ($\Delta t = 10$ ns). (b) Energy efficiency as a function of GST thickness for different pulse widths. The pulse power is selected as low as 5 mW for (a) and (b) in order to heat the aGST without inducing any phase transition. (c) GST-thickness dependent energy efficiency for crystallization ($\eta_{ac}$) actuated by a pulse with optimal power (blue line) and fixed power (red line).

**Comparison with ITO and p-i-n Heaters.** From the above analysis, the proposed PINCs with graphene heaters exhibit excellent optical and heating performance. However, there exist two other candidates for transparent heaters. First, Indium tin oxide (ITO) is a common transparent conductor that has been widely used in optoelectronics and display technology. Moreover, silicon itself could act as a transparent heater as long as the cores of the waveguides are not heavily doped. In this case, to achieve enough conductivity while maintaining low loss, a p-i-n junction could be adopted.[25] This type of heaters is only valid for silicon photonic platform in contrast to those nonvolatile silicon nitride photonic devices. Here, we compare the performance of PINCs with ITO, p-i-n, and graphene heaters through similar electro-thermal models (see Methods). As illustrated in Figure S8, all three types of PINCs have the same rib waveguides and electrodes, but for the PINCs with ITO heaters, the rib waveguides are conformally covered with 20-nm-thick ITO while for the PINCs with p-i-n heaters, the slabs are heavily doped by boron and phosphorus ion implantation ($10^{20}$ cm$^{-3}$), 100 nm away from the left and right edge of the rib, respectively.

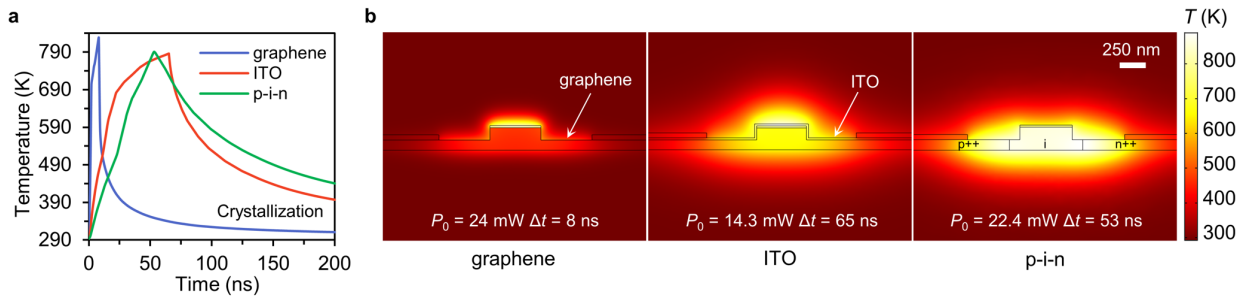

**Figure 6.** Comparison of temperature response of the PINCs with different heaters. (a) Transient temperature response of the PINCs with different heaters for crystallization. (b) Temperature distribution at the end of a pulse during the crystallization process in (a) for the PINCs with graphene, ITO, and p-i-n heaters. The power and width of the Set pulse for three heaters are listed in the corresponding figures. p++ (n++), heavily doped p (n)-type silicon region. i, intrinsic silicon region.

Due to the very different structures and material properties (Supporting Section 6), the PINCs with these heaters display extremely distinct transient temperature response and temperature distributions for optimal crystallization (Figure 6) and amorphization (Figure S9) processes. As a result, their heating performance including the switching speed and energy efficiency is also very different. For instance, the PINCs with graphene heaters show the highest switching speed thanks to the ultra-low heat capacity and high in-plane thermal conductivity of graphene, the PINCs with ITO heaters exhibit very large temperature gradient near the GST due to the low thermal conductivity of ITO, and the PINCs with p-i-n heaters have severe thermal diffusion because of the high thermal conductivity of silicon. To quantitively evaluate their heating performance, we define a heating figure of merit $FOM_2 = 1/E_{tot}/t_{ac}$, where $E_{tot}$ indicates the total energy consumption for one cycle of switching that is the summation of the optimal electrical pulse energy for crystallization ($E_{ac}$) and amorphization ($E_{ca}$) and $t_{ac}$ denotes the crystallization period that determines the switching speed. The overall figure of merit ($FOM$) for the PINCs can thus be described as the product of the optical and heating figure of merits (i.e. $FOM = FOM_1 \times FOM_2$). Table 1 lists the essential performance and figure of merits for the PINCs with three types of heaters. According to Table 1, without tuning the Fermi level, the PINCs with graphene heaters can provide the best heating performance and overall performance with $FOM_2$ and $FOM$ two orders of magnitude higher. Whereas the PINCs with p-i-n heaters have the best optical performance due to the low loss, they have the worst heating performance due to the severe heat dissipation. The overall performance for the PINCs with ITO and p-i-n heaters is close, but for ITO heaters, planarization is needed for practical applications and the large temperature gradient near the GST is concerning. In contrast, for p-i-n heaters the fabrication is relatively simple and CMOS-compatible and the temperature gradient is small due to the high thermal conductivity of silicon. In fact, such p-i-n heaters are already part of the silicon photonics foundry processes and thus can be readily adopted for phase-change material photonics. The optical performance and the conductivity of the heaters for all the PINCs strongly rely on the carrier density of the heater materials. By electrical tuning the Fermi level to the Pauli blocking region, the additional loss from graphene is significantly suppressed so the PINCs with graphene will have the best optical performance. In the meanwhile, the conductivity of graphene is desirably increased, meaning that a lower operation voltage is needed (the present voltage for Reset is ~15 V). For the PINCs with ITO and p-i-n heaters, since the increase of the carrier density will increase the extra loss, there exists a tradeoff to keep a moderate conductivity while maintaining low optical loss.

**Table 1. Performance Comparison of the PINCs with Different Heaters**

| Heater | $\Delta n_{eff}$ | $\kappa_{effa}$ | Crystallization | | | | Amorphization | | | | $FOM_1$ | $FOM_2$ (nJ⁻¹·ns⁻¹) | $FOM$ (nJ⁻¹·ns⁻¹) |
|---|---|---|---|---|---|---|---|---|---|---|---|---|---|
| | | | $\Delta t_{ac}$ (ns) | $\tau_{ac}$ (ns) | $P_0$ (mW) | $E_{ac}$ (nJ) | $\Delta t_{ca}$ (ns) | $\tau_{ca}$ (ns) | $P_0$ (mW) | $E_{ca}$ (nJ) | | | |
| graphene | 0.22 | $4.38\times10^{-3}$ | 8 | 4.47 | 24 | 0.192 | 0.22 | 1.09 | 300 | 0.066 | 50 or 140 (gated) | 0.31 | 16 or 43 (gated) |
| ITO | 0.22 | $4.74\times10^{-3}$ | 65 | 58.84 | 14.3 | 0.93 | 2.4 | 24.66 | 250 | 0.6 | 46 | 0.0053 | 0.24 |
| p-i-n | 0.21 | $1.82\times10^{-3}$ | 53 | 108.19 | 22.4 | 1.187 | 20 | 56.02 | 58.8 | 1.176 | 117 | 0.0026 | 0.31 |

$\Delta t_{ac}$ ($\Delta t_{ca}$): pulse width for crystallization (amorphization).
$\tau_{ac}$ ($\tau_{ca}$): dead time for crystallization (amorphization).
$E_{ac}$ ($E_{ca}$): optimal pulse energy for crystallization (amorphization).

## CONCLUSIONS

In conclusion, we have modeled and analyzed electrical switching of nonvolatile GST-clad integrated nanophotonic cells with graphene heaters on the programmable GST-on-silicon platform. By leveraging the ultra-low heat capacity and high in-plane thermal conductivity of graphene, a high switching speed of ~80 MHz and high energy efficiency of 19.2 aJ/nm³ (6.6 aJ/nm³) for crystallization (amorphization) are optimally achieved (via a thin film of GST and a short pulse optimized by high power) for complete phase transitions ensuring strong attenuation (~6.46 dB/μm) and optical phase (~0.28 π/μm at 1550 nm) modulation. Gigahertz operations and energy efficiency near the fundamental limit are possible for partial crystallization or amorphization during multi-level operations. Compared with ITO and silicon p-i-n heaters, the PINCs with graphene heaters have the best heating and overall performance with two orders of magnitude higher figure of merits. By gating the graphene to the Pauli blocking region, even better optical performance and lower operation voltage can be expected. To further optimize the heating performance and conduct multi-level operations, a single structured pulse or pulse sequences can be considered. The optical performance for multi-level operations can then be determined by the degree of crystallization according to the electro-thermal model.[25,45] With high speed, high energy efficiency, and small footprints while maintaining good optical performance, our proposed PINCs with graphene heaters allow scalable control over the states of PCMs and thus promise the development of the future large-scale PCM-based programmable PICs. The comprehensive model built in this work also assist the analysis and understanding of the thermal-conduction heating-enabled switching processes on PICs and facilitate the design and optimization of the PINC-based devices such as nonvolatile phase-change optical switches/modulators, directional couplers, photonic memories, and optical neurons and synapses.

METHODS

**Modeling of the PINCs with Graphene Heaters.** The electrical switching of PINCs with graphene heaters is simulated by a fully coupled electro-thermal 2D time-domain FEM model using COMSOL Multiphysics. Specifically, an electrical model (Electric Currents, Shell Interface) based on the current continuity equation is utilized to predict the current and electric potential distribution in graphene. A thermal model (Heat Transfer in Solids Interface) based on the heat transfer equation $\rho C_p \dfrac{\mathrm{d}T}{\mathrm{d}t} = \nabla \cdot (k_{th} \nabla T) + Q_e$ (where $k_{th}$ is the thermal conductivity, and $Q_e$ is the heat source) is used to predict the temperature distribution in the whole device. The two models are cross-coupled via Joule heating and the temperature-dependent material properties (Supporting Section 6).

In the electrical model, the graphene is modeled as a thin electrically conductive shell (boundary) with a thickness of 0.335 nm. The metal contacts are connected to the two sides of the graphene shell and the applied pulses are assumed to have ideal shapes.

In the thermal model, the infinite element domains are adopted for the left, right, and bottom boundary regions of the model while the convective heat flux boundary condition is used on the surface with a heat transfer coefficient of 5 W/(m²·K). Considering the relative thinness of GST and graphene and high operating temperature, thermal boundary resistance (TBR) and surface-to-surface radiation boundaries are utilized. Beside electrically s, the graphene is similarly modeled as a thin thermal conduction boundary based on the thermally thin approximation with a thickness of 0.335 nm. For simplicity without losing generality, the phase transition processes are phenomenologically modeled as that the material properties of GST are weighted sums of those in the amorphous and crystalline states in a small temperature interval of $\Delta T_m = 10$ K ($\Delta T_g = 100$ K) centered at $T_m = 888$ K ($T_g = 673$ K) of GST for melting (quenching and crystallization) with latent heat of 66.81 kJ/kg (exothermic heat of 37.22 kJ/kg) involved.[52,63] Note that $T_g$ is set to be higher than usual (~423 K) due to the increased $T_g$ at a high heating rate.[63]

The optical performance of the PINCs with graphene heaters is simulated using a frequency-domain 2D FEM wave optics model through the mode analysis (eigenvalue solver). The perfectly matched layer domains are adopted for the boundary regions of the model with the scattering boundary conditions applied to all the external boundaries. The graphene is modeled as surface current density boundaries that introduce Ohmic loss due to the optical-conductivity induced surface current.[65]

**Modeling of the PINCs with ITO and p-i-n Heaters.** Based on an identical model, the PINCs with ITO heaters are simulated except that instead of being treated as a thin film boundary, the ITO here is normally modeled as a 2D domain.

For the PINCs with p-i-n heaters, a semiconductor model (Semiconductor Interface) based on the Poisson's equation, current continuity equation, and drift-diffusion current density equations[67] is exploited to estimate the electric potential, current density, and carrier density distributions in the p-i-n junctions while the thermal model is identical to that for the PINCs with graphene heaters. In particular, the Fermi-Dirac carrier statistics and Jain-Roulston bandgap narrowing model are utilized due to the high doping level. The Arora mobility model is added to simulate the effect of phonon/lattice and impurity scattering while the Fletcher mobility model is used to describe the carrier-carrier scattering at high voltage. Trap-assisted recombination and Auger recombination for high bias are also considered in the model. The metal contacts are assumed to be ideal Ohmic and the applied pulses have ideal shapes. All the other external boundaries are electrically insulated. The carrier density distribution at semiconductor equilibrium is employed to determine the complex refractive index of the doped silicon (Supporting Section 6) for mode analysis.

ASSOCIATED CONTENT

**Supporting Information**

The Supporting Information is available free of charge on the ACS Publications website at DOI: 10.1021/acsnano.xxxxxxx.

Supporting Section 1-5 describing supplementary results for optical performance, temperature distribution analysis, transient response and speed analysis, energy efficiency analysis, and comparison of the PINCs with different heaters; Supporting Section 6 discussing material parameters used in the model (PDF)

## AUTHOR INFORMATION

**Corresponding Author**

*E-mail: arka@uw.edu.

**Author Contributions**

J.Z. and A.M. conceived the project. J.Z. performed the modeling of the devices. S.Z. and S.D. conducted the thermal conductivity modeling. P.X. helped with optical mode analysis. A.M. supervised the overall progress of the project. J.Z. wrote the manuscript with input from all the authors.

**Notes**

The authors declare no competing financial interest.

## ACKNOWLEDGMENT

The research was funded by the SRC grant 2017-IN-2743 (Fund was provided by Intel), Samsung GRO, NSF-EFRI-1640986, and AFOSR grant FA9550-17-C-0017. A.M. acknowledges support from Sloan Foundation.

**Supporting Information for**

# Modeling Electrical Switching of Nonvolatile Phase-Change Integrated Nanophotonic Structures with Graphene Heaters

*Jiajiu Zheng,[†] Shifeng Zhu,[‡] Peipeng Xu,[⊥] Scott Dunham,[†,‡] Arka Majumdar[*,†,‡]*

[†]Department of Electrical and Computer Engineering, University of Washington, Seattle, WA 98195, USA

[‡]Department of Physics, University of Washington, Seattle, WA 98195, USA

[⊥]Laboratory of Infrared Materials and Devices, Advanced Technology Research Institute, Ningbo University, Ningbo 315211, China

*E-mail: arka@uw.edu.

Number of pages: 7

Number of figures: 10

## S1. Optical performance

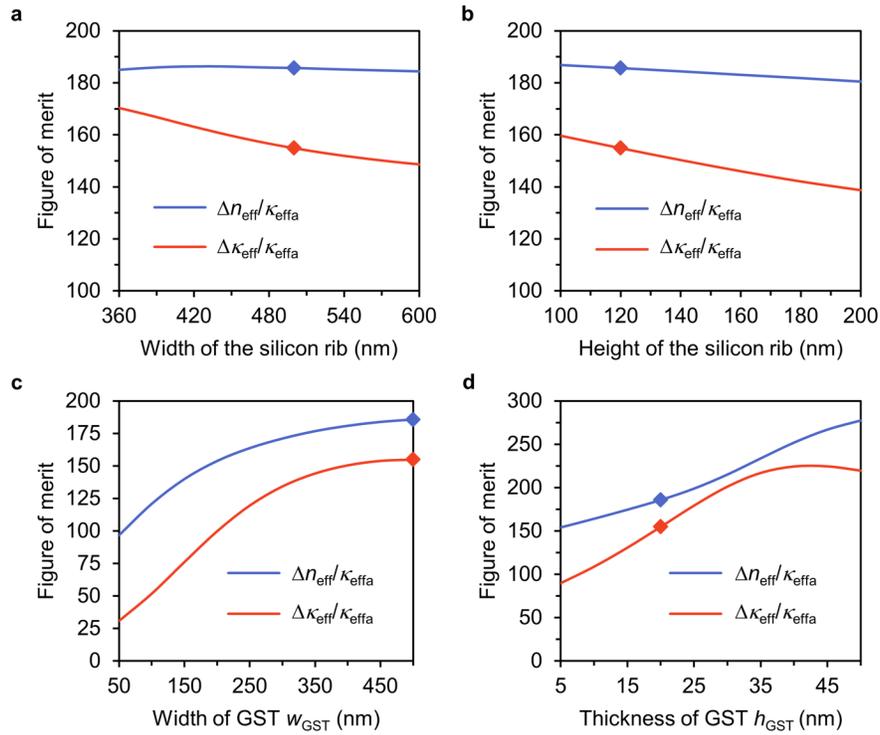

**Figure S1.** (a)-(d) Dependence of optical figure of merits on the (a) width and (b) height of the silicon rib waveguide and the (c) width and (d) thickness of the GST film at 1550 nm. By default, the width and thickness of the GST are 500 nm and 20 nm, respectively, while the width and height of the silicon rib are fixed to be 500 nm and 120 nm, respectively. This default waveguide geometry is marked as the dots in the figures and adopted for the following analysis unless specifically pointed out. Neither heaters nor metal contacts are involved at this stage.

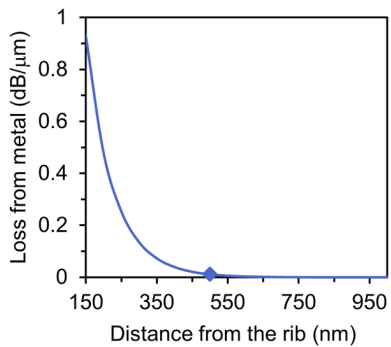

**Figure S2.** Additional loss introduced from metal contacts as a function of the distance between the edges of the electrodes and the rib waveguide. The marked dot corresponds to the adopted distance in the main text.

## S2. Temperature distribution analysis

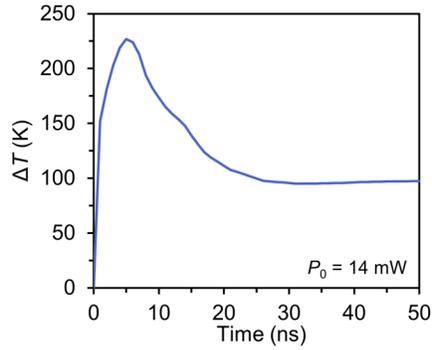

**Figure S3.** Real-time temperature gradient during the crystallization process ($P_0 = 14$ mW, $\Delta t = 50$ ns).

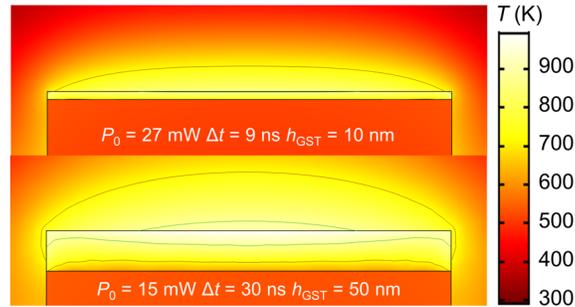

**Figure S4.** Temperature distribution at the end of a pulse indicating re-amorphization during the crystallization process due to much too high power (upper panel, $P_0 = 27$ mW, $\Delta t = 9$ ns, $h_{GST} = 10$ nm) and large GST thickness (lower panel, $P_0 = 15$ mW, $\Delta t = 30$ ns, $h_{GST} = 50$ nm). The red (green) lines denote the contour of $T = 723$ K ($T = 883$ K) for the crystallization (re-amorphization) boundary.

## S3. Transient response and speed analysis

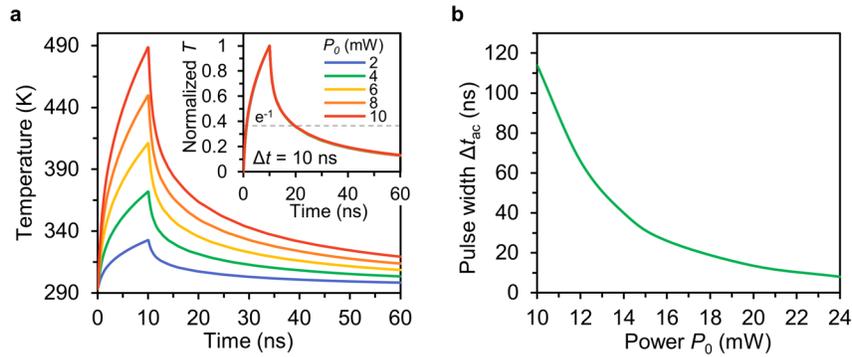

**Figure S5.** Power-dependent transient response and speed analysis. (a) Temperature response in the center of the GST cross-section to the pulses with different pulse power. Inset: corresponding normalized temperature response. The dashed line denotes the position where the temperature is 1/e of the maximum. The power of the pulse is chosen to be low enough in order to heat the cGST without inducing any phase transition and the pulse width is 10 ns. (b) Minimum pulse width required to achieve complete crystallization as a function of pulse power.

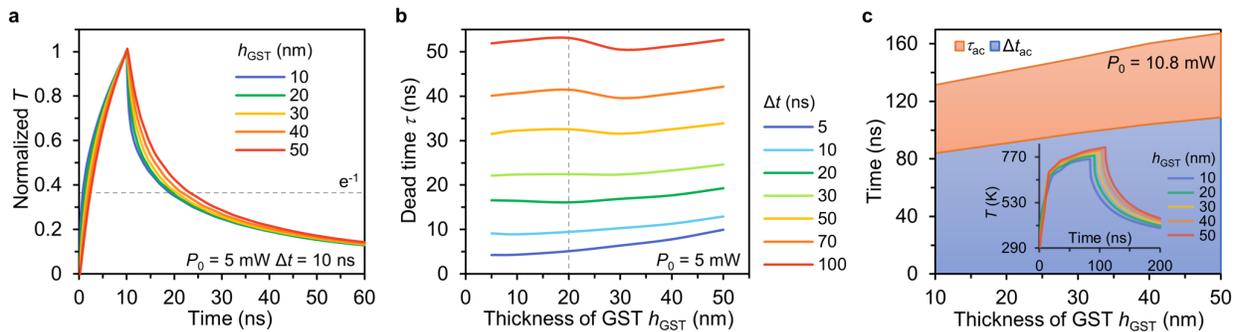

**Figure S6.** GST thickness-dependent transient response and speed analysis. (a) Normalized temperature response in the center of the GST cross-section for different GST thicknesses. The power of the pulse is chosen to be as low as 5 mW in order to heat the cGST without inducing any phase transition and the pulse width is 10 ns. The dashed line denotes the position where the temperature is 1/e of the maximum. (b) Extracted dead time as a function of GST thickness for different pulse widths. The dashed line denotes the adopted thickness in the main text. (c) Area chart of the minimum pulse width and corresponding dead time required to achieve complete crystallization actuated by a fixed pulse power ($P_0$ = 10.8 mW) as a function of GST thickness. Inset: transient temperature response of the crystallization process for different thicknesses of GST.

## S4. Energy efficiency analysis

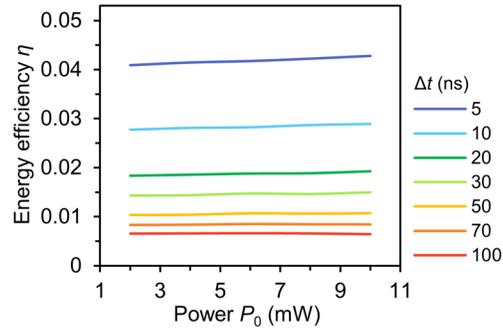

**Figure S7.** Power-dependent energy efficiency for different pulse widths. The pulse power is selected to be low enough in order to heat the aGST without inducing any phase transition.

## S5. Comparison of the PINCs with different heaters

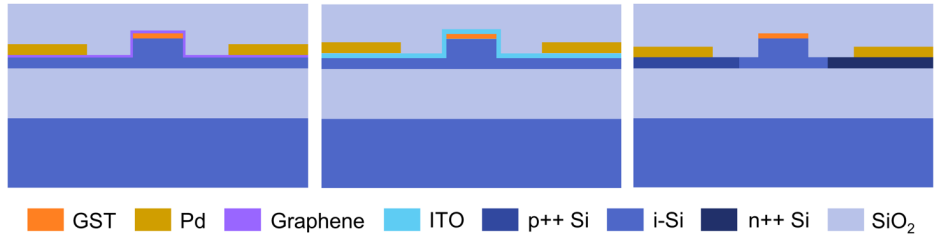

**Figure S8.** Cross-section schematics of the PINCs with graphene, ITO, and p-i-n heaters. p++ (n++), heavily doped p (n)-type silicon region. i, intrinsic silicon region.

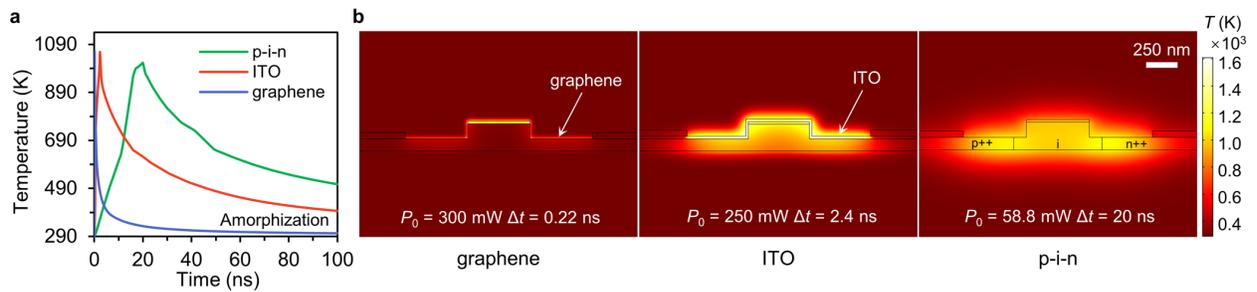

**Figure S9.** Comparison of temperature response of the PINCs with different heaters. (a) Transient temperature response of the PINCs with different heaters for amorphization. (b) Temperature distribution at the end of a pulse during the amorphization process in (a) for the PINCs with graphene, ITO, and p-i-n heaters. The power and width of the Set pulse for three heaters are listed in the corresponding figures. p++ (n++), heavily doped p (n)-type silicon region. i, intrinsic silicon region.

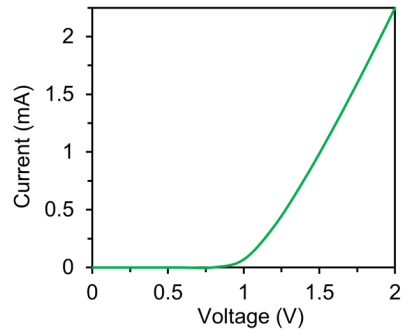

**Figure S10.** Current-voltage (I-V) curve of the silicon p-i-n diode.

## S6. Material parameters used in the model

Table S1 lists the main material parameters used in our simulations. Note that, thermal boundary resistance (TBR) was applied to all the internal boundaries and depends both on the temperature and phase of GST.[1-4] The optical conductivity of graphene is calculated according to the Kubo formula at the Fermi level of −0.26 eV.[5] The refractive index and electrical conductivity of ITO are determined by Drude model with a carrier density of $3 \times 10^{20}$ cm$^{-3}$.[6]

**Table S1. Main material parameters used in simulations.**

| | $n$ | $\kappa$ | $\sigma_{DC}$ (S m$^{-1}$) | $k$ (W m$^{-1}$ K$^{-1}$) | $C_p$ (J kg$^{-1}$ K$^{-1}$) | $\rho$ (kg m$^{-3}$) |
|---|---|---|---|---|---|---|
| **Si** | $3.4777 + \Delta n$[7] | $\Delta \kappa$[7] | N/A | $k(T)$[8] | $C_p(T)$[9] | 2329 |
| **SiO₂** | 1.444 | 0 | N/A | $k(T)$ from COMSOL | $C_p(T)$ from COMSOL | 2200 |
| **ITO** | 1.4497[6] | 0.0923[6] | $2 \times 10^5$[6] | 3.2[4] | $C_p(T)$ from COMSOL | 7100 |
| **Graphene** | $\sigma = 6.05 \times 10^{-5} + 6.19 \times 10^{-6}$ i[5,10] | | $\sigma_{DC}(T)$ ($R_s = 800$ $\Omega/\square$ at 293 K)[11] | $k(T)$ (160 at 293 K)[12-13] | $C_p(T)$[12] | 2271 |
| **Pd** | 3.1640 | 8.2121 | N/A | 71.8 | 244 | 12023 |
| **aGST** | 3.8884[14] | 0.024694[14] | N/A | 0.19[15] | 213[16] | 5870[17] |
| **cGST** | 6.6308[14] | 1.0888[14] | N/A | $k(T)$[15] | 199[16] | 6270[17] |

$n$, refractive index. $\kappa$, extinction coefficient. $\sigma$, optical conductivity. All the optical parameters are for 1,550 nm. $\sigma_{DC}$, electrical conductivity. $R_s$, sheet resistance. $k$, thermal conductivity. $C_p$, heat capacity at constant pressure. $\rho$, density. $T$, temperature. $\Delta n$ and $\Delta \kappa$ are dependent on carrier densities of electrons and holes and can be calculated from Eq. 6.3 and Eq. 6.4 in Ref. 6.

REFERENCES of MAIN TEXT: